\documentclass[fleqn,usenatbib]{mnras}
\usepackage[english]{babel}

\usepackage[T1]{fontenc}
\usepackage{ae,aecompl}

\usepackage{fancyhdr}
\usepackage{amsfonts}
\usepackage{amsmath}
\usepackage{amssymb}
\usepackage{multicol}
\usepackage{layout}
\usepackage{graphicx}
\usepackage{multirow}

\usepackage{color}
\usepackage{multirow}
\usepackage[utf8]{inputenc}
\usepackage{times}
\usepackage{natbib}
\def\be{\begin{equation}}
\def\ee{\end{equation}}

\newif\ifAMStwofonts
\AMStwofontstrue

\graphicspath{{./fig/}}

\title[agegraphic dark energy and growth of perturbations]{Agegraphic dark energy: growth index and cosmological implications}

\author[M. Malekjani et~al.]{M. Malekjani$^{1}$
\thanks{malekjani@basu.ac.ir}, S. Basilakos$^2$, A. Mehrabi$^{1}$,
Z. Davari$^{1}$ and M. Rezaei$^1$\\
$^1$ Department of Physics, Bu Ali Sina University, Hamedan, Iran\\
$2$ Academy of Athens, Research Center for Astronomy and Applied
Mathematics, Soranou Efessiou 4, 11-527 Athens, Greece }

\date{Accepted ?, Received ?; in original form \today}

\pagerange{\pageref{firstpage}--\pageref{lastpage}} \pubyear{2015}

\begin{document}

\label{firstpage}

\maketitle

\begin{abstract}
We study the main cosmological properties of the agegraphic dark
energy model at the expansion and perturbation levels. Initially,
using the latest cosmological data we implement a joint likelihood
analysis in order to constrain the cosmological parameters. Then we
test the performance of the agegraphic dark energy model at the
perturbation level and we define its difference from the usual
$\Lambda$CDM model. Within this context, we verify that the growth
index of matter fluctuations depends on the choice of the considered
agegraphic dark energy (homogeneous or clustered). In particular,
assuming a homogeneous agegraphic dark energy we find, for the first
time, that the asymptotic value of the growth index is $\gamma
\approx 5/9$, which is close to that of the usual $\Lambda$ cosmology,
$\gamma^{(\Lambda)} \approx 6/11$. Finally, if the distribution of
dark energy is clustered then we obtain $\gamma \approx 1/2$
which is $\sim 8\%$ smaller than that of the $\Lambda$CDM model.

\end{abstract}


\section{Introduction}
%%%%%%%%%%%%%%%%%%%%%%%%%%%%%%%%%%%%%%%%%%%%%%%%%%%%%%%%%%%%%%%
  The analysis of various observational data including those of
supernovae type Ia (SNIa)
\citep{Riess1998,Perlmutter1999,Kowalski2008}, cosmic microwave
background (CMB)
\citep{Komatsu2009,Komatsu2011,Jarosik:2010iu,Planck2015_XIII},
large scale structure and baryonic acoustic oscillation (BAO)
\citep{Percival2010,Tegmark:2003ud,Cole:2005sx,Eisenstein:2005su,Reid:2012sw,Blake:2011rj},
high $z$ galaxies \citep{Alcaniz:2003qy}, high $z$ galaxy clusters
\citep{Allen:2004cd,Wang1998}, weak gravitational lensing
\citep{Benjamin:2007ys,Amendola:2007rr,Fu:2007qq} strongly suggest
an accelerated expansion of the universe. In the context of General
Relativity (GR) the so-called dark energy (hereafter DE), which has
a negative pressure, is required in order to interpret the cosmic
acceleration. It is interesting to mention that from the overall
energy density, only $\sim 30\%$ consists of matter (luminous and
dark) while the rest corresponds to DE (see the analysis of
Planck2015 \citep{Planck2015_XIII} and references therein).

From the theoretical perspective, over the last two decades, a large
family of phenomenological models has been proposed to study the
cosmological features of DE. The simplest DE model is the
concordance $\rm \Lambda$ cosmology for which the equation of state
(EoS) $ w_{\rm \Lambda}$ is strictly equal to -1. Despite the fact
that the $\Lambda$CDM model fits extremely well the cosmological
data it suffers from the fine tuning and the cosmic coincidence
problems
\citep{Weinberg1989,Sahni:1999gb,Carroll2001,Padmanabhan2003,Copeland:2006wr}.
Since the nature of DE has yet unknown, different versions of
dynamical DE models have been introduced in order to alleviate the
above cosmological issues. Generally speaking, it has been proposed
that we cannot entirely understand the nature of dark energy before
the establishment of a complete theory of quantum gravity
\citep{Witten:2000zk}. Nevertheless, it is promising that
holographic dark energy (HDE) models \citep{Horava2000,Thomas2002}
inspired by the principles of quantum gravity can be suggested and
may hopefully provide an efficient explanation for the dynamical
nature of DE. Specifically, the holographic principle
\citep{Susskind1995} points out that in a finite-size physical
system the number of degrees of freedom should be finite and bounded
by the area of its boundary \citep{Cohen1999}. In other words, the
total energy of a physical system with size $L$ obeys the following
inequality $L^3\rho_{\Lambda}\leq Lm_{\rm p}^2$, where $\rho_{\rm
\Lambda}$ is the quantum zero-point energy density and $m_{\rm
p}=1/\sqrt{8 \pi G}$ is the Planck mass. Applying the latter
arguments to cosmological scales it has been found \citep[see
][]{Li2004} that the density of the HDE is given by
\begin{equation}\label{eq:density}
 \rho_{\rm d}=3n^2m_{\rm p}^2L^{-2}\;,
 \end{equation}
where $n$ is a positive constant and the coefficient $3$ is used for
convenience.
%In this model, it has been shown that the cosmic coincidence
%problem can be alleviated by inflation \citep{Li2004}.
Obviously, in this case the features of DE strongly depend on the
definition of the size $L$ in equation (\ref{eq:density}). If we
assume $L$ to be the Hubble radius $H^{-1}$ then we cannot produce
an accelerated expansion of the universe
\citep{Horava2000,Cataldo2001,Thomas2002,Hsu2004}. Another choice
would be to replace $L$ with the particle horizon but again we would
not be able to extract cosmic acceleration
\citep{Horava2000,Cataldo2001,Thomas2002,Hsu2004}. The final choice
for $L$ is to use the event horizon (first introduced by
\cite{Li2004}). In this case, the HDE model is able to provide
cosmic acceleration and it is consistent with observations
\citep{Pavon2005,Zimdahl2007,Sheykhi:2011cn}. Notice, that the HDE
model has been widely investigated in the literature
\citep{Huang:2004wt,Huang2004b,Gong2004,Gong2005a,Gong2005b,Zhang:2005hs,Zhang:2007sh,
Elizalde:2005ju,Guberina:2005fb,Guberina:2006qh,BeltranAlmeida:2006is,Wang:2004nqa,Shen:2004ck}.

Since the HDE model is obtained by choosing the event horizon length
scale, an obvious drawback concerning causality appears in this
scenario. Recently, a new DE model, dubbed agegraphic dark energy
(ADE) model, has been suggested by \cite{Cai2007} in order to
alleviate the above problem. In particular, combining the
uncertainty principle in quantum mechanics and the gravitational
effects of GR Karolyhazy and his collaborators
\citep{Karolyhazy:1966zz,Karolyhazy1982,Karolyhazy1986} made an
interesting observation concerning the distance measurement for the
Minkowski spacetime through a light-clock Gedanken experiment
\citep[see also][]{Maziashvili:2007zz}. They found that the distance
$t$ in Minkowski spacetime cannot be known to a better accuracy than
$\delta t=\beta t_{\rm p}^{2/3}t^{1/3}$, where $\beta$ is a
dimensionless constant of order ${\cal O}(1)$ \citep[see
also][]{Maziashvili:2007zz}. \footnote{Through out this work we use
the units $\hbar=c=k_{\rm B}=1$. Hence we have $l_{\rm p}=t_{\rm
p}=1/m_{\rm p}$ ,where $l_{\rm p}$, $t_{\rm p}$ and $m_{\rm p}$ are
the reduced Planck length, time and mass, respectively.} Based on
the Karolyhazy relation, \cite{Maziashvili:2007zz} argued that the
energy density of metric fluctuations in the Minkowski spacetime is
written as \citep[see also][]{Maziashvili:2007dk}

\begin{equation}\label{eq:energy2}
\rho_{d}\sim\frac{1}{t_{p}^{2}t^{2}}\sim \frac{m_{p}}{t^{2}}\;,
\end{equation}
where $ m_{p}$ and $t_{p}$ are the reduced Plank mass and the Plank
time, respectively \citep[see
also][]{Sasakura:1999xp,Ng:1993jb,Ng:1995km,Krauss:2004fb,Christiansen:2005yg,Arzano:2006wp,Ng:2007bp}.
Using equation (\ref{eq:energy2}) \cite{Cai2007} proposed another
version of holographic DE the so-called agegraphic dark energy (ADE)
in which the time scale $t$ is chosen to be equal with the age of
the universe $T= \int^{t}_0dt=\int_{0}^{a}\frac{da}{aH} $, with $a$
the scale factor of the universe and $H$ the Hubble parameter.
Therefore, the ADE energy density is given by \citep{Cai2007}
 \begin{equation}
 \rho_{d}=\frac{3n^{2}m_{p}^{2}}{T^{2}}\label{eq:energy3}
 \end{equation}
where $n$ is a free parameter and the coefficient $3$ appears for
convenience. The present value of the age of universe ($T_{\rm
0}\sim H_{\rm 0}^{-1}$) implies that $n$ is of order ${\cal O}(1)$.
It has been shown that the condition $n>1$ is required in order to
have cosmic acceleration \citep{Cai2007}. Although, the ADE scenario
does not suffer from the causality problem \citep{Cai2007} it faces
some problems towards describing the matter-dominated epoch
\citep{Wei:2007ty,Neupane:2007fw,Wei:2007xu}. To overcome this issue
\cite{Wei:2007ty} proposed a new agegraphic dark energy (NADE)
model, in which the cosmic time $t$ is replaced by the conformal
time $ \eta=
\int^{t}_0{\frac{dt}{a(t)}}=\int^{a}_0{\frac{da}{a^{2}H}}$ and thus
the energy density in this case becomes \citep{Wei:2007ty}

\begin{equation}
 \rho_{d}=\frac{3n^{2}m_{p}^{2}}{\eta^{2}}\label{roo} \;.
 \end{equation}
It is interesting to mention that \cite{Kim:2007iv} showed that the
NADE model provides the proper matter-dominated and
radiation-dominated epochs, in the case of $n>2.68$ and $n>2.51$,
respectively. Also \cite{Wei:2007xu} found that the coincidence
problem can be alleviated naturally in this model and using the
cosmological data (SNIa, CMB etc) they obtained
$n=2.716^{+0.111}_{-0.109}$. We would like to point out that the
cosmological properties of the ADE and the NADE models can be found
in
\cite{Kim:2008hz,Setare:2010zy,Karami:2010qe,Sheykhi:2010jn,Sheykhi:2009rk,Sheykhi:2009yn,
Sheykhi:2009sz,Sheykhi:2009sn,
Lee:2008zzw,Jawad:2014ssa,Liu:2012kha,
Zhang:2012pr,Farajollahi:2012zz,Zhai:2011pp,
Chen:2011rz,Sun:2011vg,Lemets:2010qz,Zhang:2010im,Liu:2010ci,Karami:2010aq,Malekjani:2010uv,Jamil:2010vr,
Karami:2009wd,Sheykhi:2009yn,Sheykhi:2009sz,Sheykhi:2009rk,Wu:2008jt,Zhang:2008mb,Kim:2007iv,Neupane:2007ra,Wei:2007xu}.\\

%In this work we study the NADE model at the perturbation level.
Furthermore, it is well known that beyond the expansion rate of the
universe, DE affects the formation of cosmic structures
\citep{Peebles1993,Tegmark:2003ud}. Usually, in dynamical DE models
with $w_{\rm d}\neq-1$, one can assume that the DE perturbations
behave in a similar fashion to matter
 \citep{Abramo2007,Abramo:2008ip,Abramo2009a,Batista:2013oca,Batista:2014uoa,Mehrabi:2014ema,Malekjani:2015pza,Mehrabi:2015hva}.
In principle, the effective sound speed $c_{\rm eff}^{2}=\delta
p_{\rm d}/\delta\rho_{\rm d}$ is introduced in order to describe the
DE clustering. In particular, if DE is homogeneous then we have
$c^{2}_{\rm eff}=1$, while for clustered DE models we use
$c^{2}_{\rm eff}=0$. In the homogeneous case, the sound horizon of
DE is close to the Hubble length. This does not hold for $c^{2}_{\rm
eff}=0$ (clustered DE) which implies that the perturbations of DE
grow, via gravitational instability, in sub-Hubble scales
\citep{ArmendarizPicon:1999rj,ArmendarizPicon:2000dh,Garriga:1999vw,Akhoury:2011hr}.
 The growth of matter perturbations in cosmologies where DE is
allowed to have clustering has been widely investigated in the
literature
\citep{Batista:2013oca,Erickson:2001bq,Bean:2003fb,Hu:2004yd,Ballesteros:2008qk,
Basilakos:2009mz,dePutter:2010vy,Sapone:2012nh,Dossett:2013npa,Basse:2013zua,
Batista:2014uoa,Pace:2014taa,Steigerwald:2014ava,Mehrabi:2014ema,Mehrabi:2015hva,Malekjani:2015pza,Basilakos:2014yda,Nesseris:2014mfa}.

In this article, following the lines of the above studies, we
attempt to investigate the NADE model at the background and
perturbation levels. Specifically, we organize the manuscript as
follows: in section (2) we start with a brief presentation of the
NADE model and in section (3) we investigate the growth of matter
perturbations. In section (4), using the latest cosmological data
and the growth rate data, we perform a joint likelihood analysis in
order to constraint the free parameters of the model. In section (5)
we discuss the growth index and finally, we summarize our results in
section (6).

%************************************************************************
%***********************************************************************
\section{Background evolution in NADE cosmology}
 Considering a spatially flat Friedmann-Robertson-Walker (FRW)
metric, one can show that the Hubble parameter takes the following
form
\begin{equation}
 H^{2}=\frac{1}{3m_{p}^{2}}( \rho_{d}+ \rho_{m})\;, \label{frid}
\end{equation}
where $\rho_{m}$ and $\rho_{d}$ are the energy densities of
pressure-less matter and DE respectively. In the case of
non-interacting DE models the corresponding densities satisfy the
following continuity equations
\begin{eqnarray}
&&\dot{\rho}_{d}+3H\rho_{d}(1+w_{d})=0 \label{contd}\\
&&\dot{\rho}_{m}+3H\rho_{m}=0,\label{contm}
\end{eqnarray}
where $\rho_{m}\propto a^{-3}$ and $\rho_{d}$ is given by equation
(\ref{roo}). Therefore, introducing the density parameter $
\Omega_{d} \equiv\frac{\rho_{\rm d}}{3m_{p}^{2}H^{2} }$, we obtain
\begin{equation}\label{eq:eta1}
\Omega_{d}=\frac{n^{2}}{H^{2}\eta^{2}} .
\end{equation}
Of course, one can easily show that equation (\ref{frid}) is reduced to
$\Omega_{d}+\Omega_{m}=1$, where $ \Omega_{\rm m}
\equiv\frac{\rho_{\rm m}}{3m_{p}^{2}H^{2} }$.

Furthermore, differentiating with respect to cosmic time
equation (\ref{roo}) and using
equations (\ref{contd}) and (\ref{eq:eta1}), we derive
the corresponding equation of state (EoS) parameter
\begin{equation}
w_{d}=-1+\frac{2}{3na}\sqrt{\Omega_{d}},\label{wd}
\end{equation}
where $a(z)=1/(1+z)$ is the scale factor of the universe. Evidently,
the above EoS parameter obeys the inequality $w_{d}>-1$ and thus it
cannot enter the phantom regime. As expected, at late enough times
($a\rightarrow \infty$) NADE tends to $\Lambda$CDM
($w_{d}\rightarrow -1$). On the other hand, in the matter-dominated
epoch, we have $H^{2}\propto\rho_{m}\propto a^{-3}$, which implies
$a^{\frac{1}{2}} da\propto dt=a d\eta$. Hence, $\eta\propto
a^{\frac{1}{2}}$ and using equations (\ref{roo}), (\ref{eq:eta1})
and (\ref{contd}) we easily find $\rho_{d}\propto a^{-1}$,
$\Omega_{d}\simeq n^{2}a^{2}/4$ \citep{Wei:2007xu} and thus $w_{\rm
d}\simeq -2/3$. \footnote{In the radiation dominated era we have
$H^2\propto\rho_{r}\propto a^{-4}$ and $\eta\propto a$. In this case
one can prove that $\Omega_{d} \simeq n^{2}a^{2}$ and $w_{\rm
d}\simeq -1/3$ \citep{Wei:2007xu}.} Also, following the notations of
\cite{Wei:2007ty} one can show \citep{Wei:2007ty}
\begin{equation}
\frac{d\Omega_{\rm
d}}{da}=\frac{\Omega_{d}}{a}(1-\Omega_{d})(3-\frac{2}{na}\sqrt{\Omega_{d}}).\label{omega}
\end{equation}
 Utilizing the Friedmann equation (\ref{frid}), the continuity
equations (\ref{contd},\ref{contm}) and the dimensionless Hubble
parameter $E(a)=H(a)/H_{0}$, we have\footnote{For the $\Lambda$CDM
model we utilize
$E(a)=\sqrt{\Omega_{m0}a^{-3}+\Omega_{r0}a^{-4}+\Omega_{d0}}$.}
%, in a flat geomety can be derived as follows
\begin{equation}
E^{2}(a)=\frac{\Omega_{m0}a^{-3}+\Omega_{r0}a^{-4}}{1-\Omega_{d}(a)}\;,\label{Ez}
\end{equation}
where $\Omega_{\rm d}$ satisfies equation (\ref{omega}). We proceed
now by solving the system of coupled equations
(\ref{wd},\ref{omega},\ref{Ez}) in order to compute the evolution of
the $H(z)$, $w(z)$ and $\Omega_{d}(z)$. As far as the initial
conditions are concerned, we start from  $a_{\rm i}=0.0005$ which is
deep enough in the matter-dominated era \citep[see
also][]{Wei:2007xu}. Consequently, the initial value of the DE
density parameter is given by $\Omega_{\rm d}(a_{\rm i})=n^{2}a_{\rm
i}^{2}/4$.

\begin{figure}
 \centering
 \includegraphics[width=8cm]{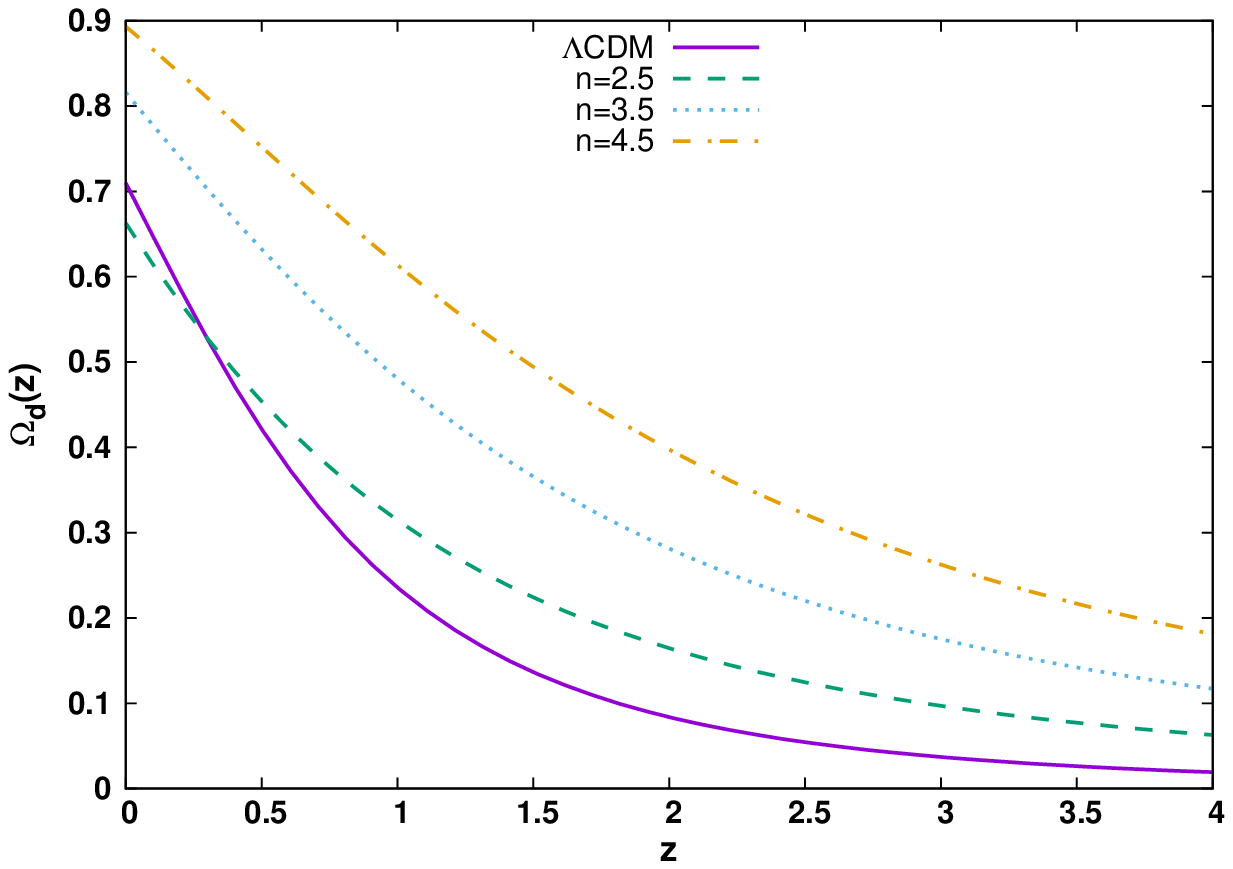}
 \includegraphics[width=8cm]{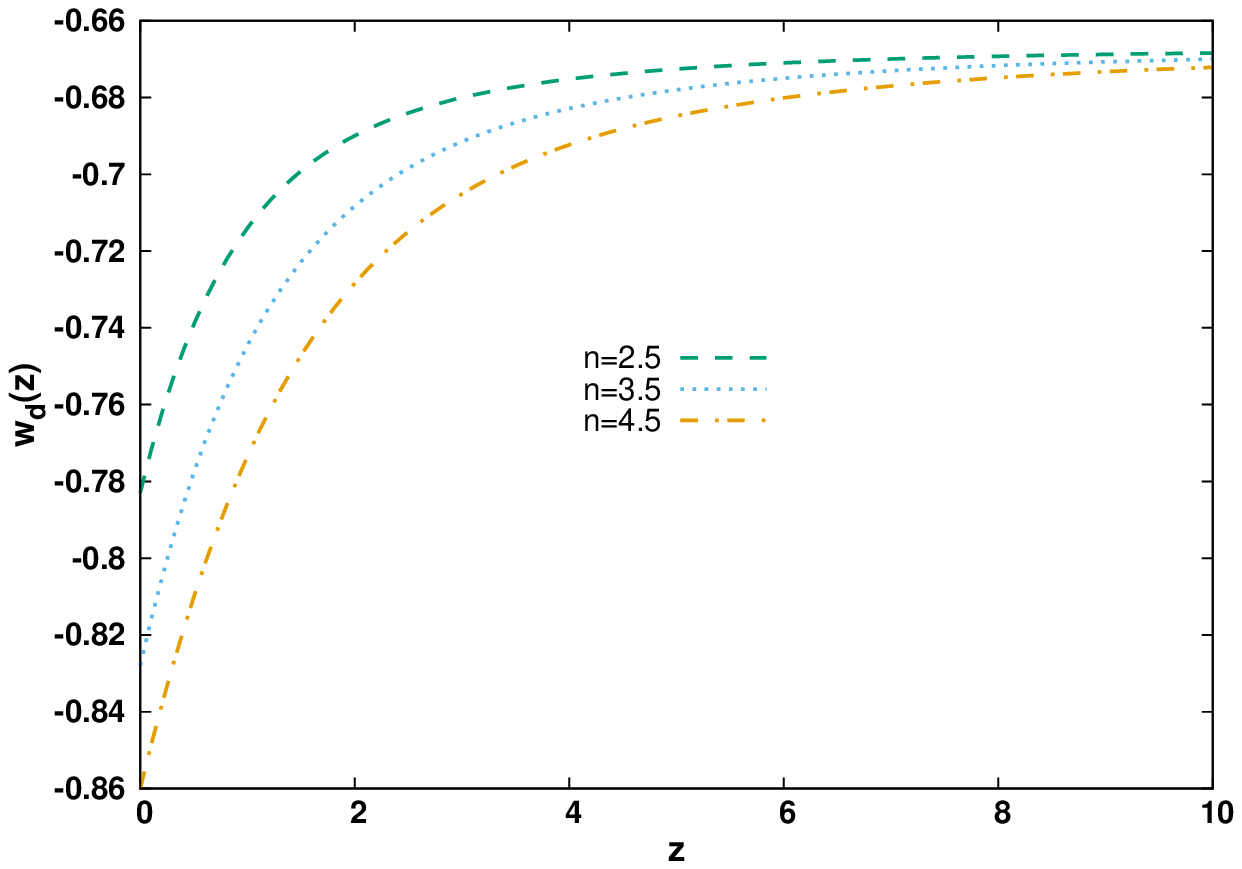}
 \includegraphics[width=8cm]{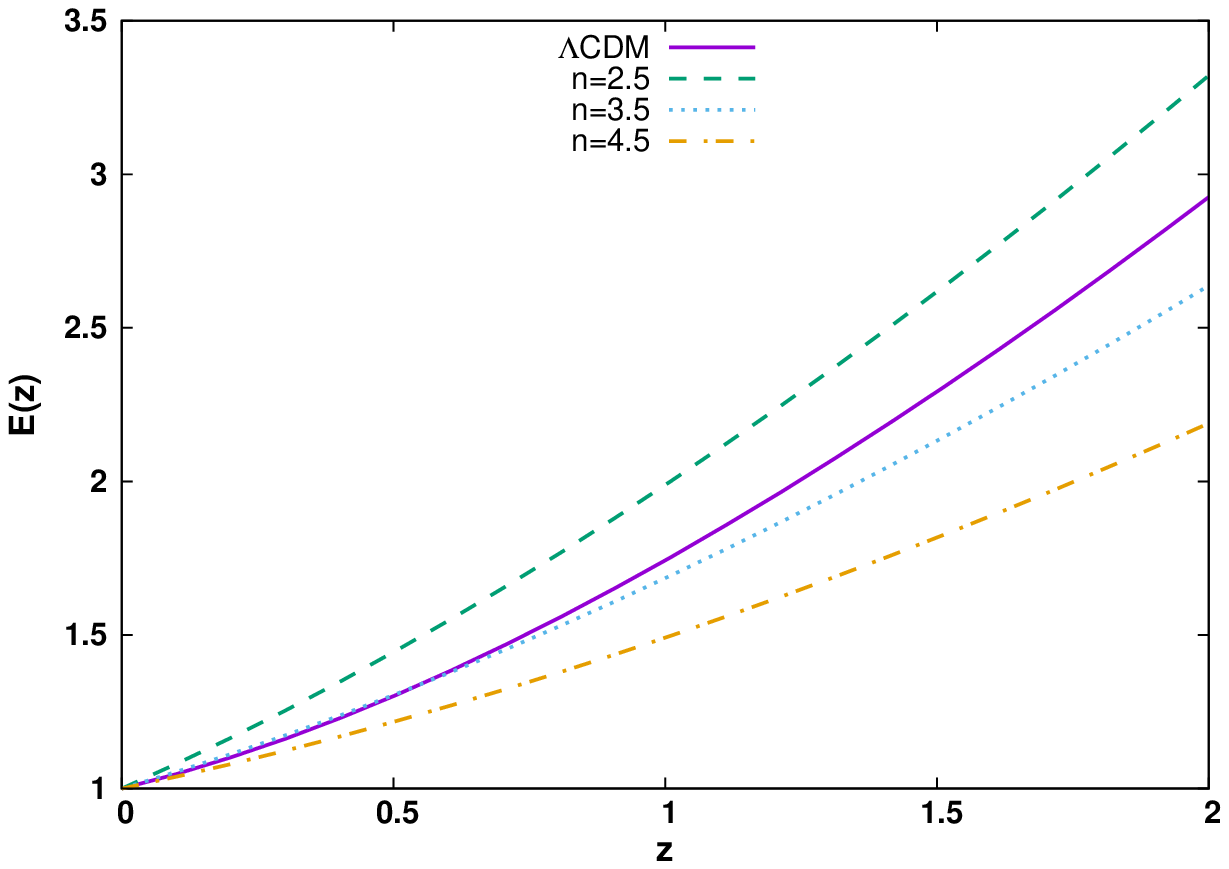}
 \caption{The redshift evolution of $\Omega_{d}$ (top panel),
$w_{d}$ (middle panel) and $E=H/H_{0}$ (bottom panel). The dashed,
dotted and dotted dashed curves correspond to NADE models with $n =
2.5$, $n = 3.5$ and $n = 4.5$, respectively. Notice, that the
reference $\Lambda$CDM model is shown by solid curve.}
 \label{figa}
\end{figure}

In Fig.(\ref{figa}) we show the redshift evolution of
$\Omega_{d}(z)$ (top panel), $w_{d}(z)$ (middle panel) and $E(z)$
(bottom panel) for different values of the model parameter $n=2.5$
(dashed line), $3.5$ (dotted line) and $4.5$ (dotted dashed line). The
concordance $\Lambda$CDM cosmology is also plotted for comparison
(see solid line). As expected, the aforementioned cosmological
quantities depend on the choice of $n$. Overall, we observe that the
EoS parameter remains in the quintessence regime and it lies in the
interval $-1 \leq w_{d} \leq -\frac{2}{3}$. Also, for large values
of $n$, the EoS parameter tends  to -1 at the present epoch. The
evolution of the DE density parameter shows that for large values of
$n>3$, $\Omega_{\rm d}$ is large with respect to that of the
concordance $\Lambda$ cosmology. Regarding the normalized Hubble
parameter $E(z)$ we see that in the case of $n>3$ we have $E_{\rm
NADE}(z)<E_{\Lambda}(z)$, while the opposite holds for $n=2.5$.

\section{growth of perturbations}
In this section we proceed with the study of the linear growth of
matter perturbations in the NADE model. An important ingredient in
this kind of studies is that one can use the so called
pseudo-Newtonian approach \citep{Abramo2007,Pace2010}. It has been
shown that in sub-Hubble scales the results of pseudo-Newtonian
dynamics are consistent with those of GR \citep{Abramo2007}.

According to \cite{Abramo:2008ip}, the evolution of matter
and DE perturbations can be
described from the following equations:

\begin{eqnarray}
&&\dot{\delta}_{m}+\frac{\theta_{m}}{a}=0\label{dm}\\
&&\dot{\delta}_{d}+(1+w_{d})\frac{\theta_{d}}{a}+3H(c_{\rm eff}^{2}-w_{d})\delta_{d}=0\label{de}\\
&&\dot{\theta}_{m}+H\theta_{m}-\frac{k^{2}\phi}{a}=0\label{tm}\\
&&\dot{\theta}_{d}+H\theta_{d}-\frac{k^{2}c_{\rm eff}^{2}\theta_{d}}{(1+w_{d})a}-\frac{k^{2}\phi}{a}=0\label{td}
\end{eqnarray}
where dot means derivative with respect to cosmic time $t$, $k$ is
the wave number of perturbations and $c_{\rm eff}^{2}$ is the
effective sound speed. Moreover, the amount of DE clustering
strongly depends on the choice of $c_{\rm eff}^2$. Indeed, for
$c_{\rm eff}^2=0$, the perturbations of DE can grow in a similar
fashion to matter perturbations \citep[see
also][]{Abramo:2008ip,Batista:2013oca,Batista:2014uoa}. For this
scenario, the fact that the perturbations of DE are affected by the
negative pressure implies that the amplitude of DE perturbations is
small with respect to that of dark matter fluctuations. On the other
hand, for the homogeneous case we have $c_{\rm eff}^2=1$ which means
that DE perturbations are vanished in sub-Hubble scales.

Also, using the Poisson equation in sub-Hubble scales and with the
aid of the above equations we have
\begin{equation}\label{pois}
-\frac{k^{2}}{a^{2}}\phi=\frac{3}{2}H^{2}[\Omega_{m}\delta_{m}+(1+3c_{\rm
eff}^{2})\Omega_{d}\delta_{d}].
\end{equation}

Replacing equation (\ref{pois}) with equations (\ref{td}) \&
(\ref{tm}), eliminating $\theta_{\rm m}$ and $\theta_{\rm d}$ from
the system of equations (\ref{dm},\ref{de},\ref{tm},\ref{td}) and
changing the variables for $t$ to the scale factor $a(t)$ we find
\begin{eqnarray}
&&\delta_{m}^{''}+A_{m}\delta_{m}^{'}+B_{m}\delta_{m}=\frac{3}{2a^{2}}(\Omega_{m}\delta_{m}+\Omega_{d}\delta_{d}),\label{ddm}\\
&&\delta_{d}^{''}+A_{d}\delta_{d}^{'}+B_{d}\delta_{d}=\frac{3}{2a^{2}}(1+w_{d})(\Omega_{m}\delta_{m}+\Omega_{d}\delta_{d}),\label{dd}
\end{eqnarray}
where we have set $c_{\rm eff}^2=0$
and prime denotes derivative with respect
to scale factor. Moreover, the corresponding
coefficients are written as
\begin{eqnarray}
&& A_{m}=\frac{3}{2a}(1-w_{d}\Omega_{d});\nonumber\\
&& B_{m}=0;\nonumber\\
&& A_{d}=\frac{1}{a}[-3w_{d}-\frac{aw_{d}^{'}}{1+w_{d}}+\frac{3}{2}(1-w_{d}\Omega_{d})];\nonumber\\
&& B_{d}=\frac{1}{a^{2}}[-aw_{d}^{'}+\frac{aw_{d}^{'}w_{d}}{1+w_d}-\frac{1}{2}w_{d}(1-3w_{d}\Omega_{d})].\nonumber\\
\end{eqnarray}
It is important to point out that the growth of DE
perturbations strongly depends on the evolution of EoS parameter $w_{\rm d}$.
On the other hand, the growth of matter perturbations
are directly affected by the perturbations of DE through
the system of equations (\ref{ddm}) \& (\ref{dd}).

 Now, we numerically solve the system of Eqs.(\ref{ddm}) and
(\ref{dd}), using simultaneously the background equations
(\ref{wd}), (\ref{omega}) and (\ref{Ez}). Concerning the initial
conditions we impose the following restriction: at $z_{\rm i}=2000$
($a_{\rm i}=0.0005$) we use $\delta_{\rm mi}(z_{\rm i})=8 \times
10^{-5}$. Additionally, we also adopt the initial conditions
provided by \cite{Batista:2013oca} and \cite{Mehrabi:2015kta} as
follows

\begin{eqnarray}
&&\delta_{\rm mi}^{\prime}=\frac{\delta_{\rm mi}}{a_{\rm i}}\;, \label{IC1}\\
&&\delta_{\rm di}=\frac{1+w_{\rm di}}{1-3w_{\rm di}}\delta_{\rm mi}\;, \label{IC2}\\
&&\delta_{\rm di}^{\prime}=\frac{4w_{\rm di}^{\prime}}{(1-3w_{\rm di})^2}\delta_{\rm mi}+\frac{1+w_{\rm di}}{1-3w_{\rm di}}\delta_{\rm mi}^{\prime}\; \label{IC3},
\end{eqnarray}
where $w_{\rm di}=w_{d}(z_{i})$.
%is the value of EoS parameter of NADE model at the initial scale factor $a_{\rm i}=0.0005$.

As expected, in the case of homogeneous NADE model ( $\delta_{\rm
d}\equiv 0$), equation (\ref{ddm}) is reduced to the well known
differential equation \citep[see also][]{Pace2010,Mehrabi:2014ema}
\begin{equation}
\delta_{m}^{''}+\frac{3}{2a}(1-w_{\rm d}\Omega_{\rm d})\delta_{m}^{'}-\frac{3\Omega_{m0}}{2a^{5}E^2}\delta_{m}=0,\label{ddm2} \;.
\end{equation}
Therefore, if we allow NADE to have clustering then we solve the
system of equations (\ref{ddm}) and (\ref{dd}) with the aid of the
initial conditions (\ref{IC1}, \ref{IC2} and \ref{IC3}). On the
other hand, in the case of homogeneous NADE model we only solve
equation (\ref{ddm2}). Once the matter perturbation $\delta_{\rm m}$
and DE perturbation $\delta_{\rm d}$ are obtained, it is relatively
easy to compute the linear growth factor scaled to unity at the
present time $D(a)=\delta_{\rm m}(a)/\delta_{\rm m}(a=1)$.

\begin{figure}
 \centering
 \includegraphics[width=8cm]{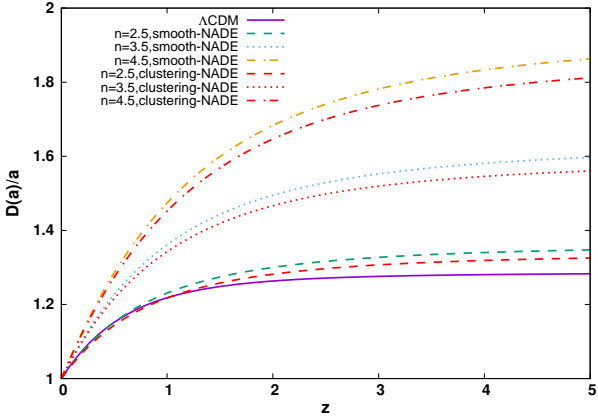}
 \caption{The redshift evolution of $D(a)/a$. Notice, that
the growth factor is normalized to unity at the present time. The
purple solid curve shows the concordance $\Lambda$CDM model. In the
case of smooth NADE model the corresponding lines are the same to
those of figure (\ref{figa}). Red- dashed, red- dotted and
red-dotted-dashed curves represent the clustering NADE models with
$n=2.5$, $n=3.5$ and $n=4.5$, respectively.}
 \label{figb}
\end{figure}

In figure (\ref{figb}), we show the quantity $D(a)/a$ as a function
of redshift, $a(z)=1/(1+z)$. As expected, in the Einstein de-Sitter
(EdS) model ($\Omega_{\rm m}=1$) the function $D(a)/a$ is always
equal to unity. For comparison we also plot $D(a)/a$ for the
$\Lambda$CDM cosmology (purple solid curve). Regarding the NADE
cosmological model, we plot the evolution of $D(a)/a$ for the
following two cases: smooth NADE model ($c_{\rm eff}^{2}=1$) and
clustered NADE model ($c_{\rm eff}^{2}=0$), where the model
parameter $n$ takes the values $n=2.5,3.5$ and $4.5$. The curve
styles and the corresponding colors are mentioned in the caption of
figure (\ref{figb}). In general, we find that the amplitude of
$D(a)/a$ increases as a function of $n$. We observe that at high $z$
the amplitude of the linear growth of matter fluctuations is larger
than the standard $\Lambda$CDM model. Moreover, at high redshifts
($z>2$) $D_{\Lambda}(a)/a$ reaches a plateau which implies that the
impact of the cosmological constant on the growth of cosmic
structures is practically negligible. However, this is not the case
for the NADE cosmological models, namely $D(a)/a$ seems to evolve
even at $z\sim 4$. This behavior of the growth factor at high $z$
can be easily interpreted as a small but non-negligible effect of
the DE component on the growth of perturbations as we can see in
top-panel of figure (\ref{figa}) \citep[see
also][]{Batista:2013oca,Malekjani:2015pza}. Note that at the same
redshift range and for $n=4.5$ we find that the quantity $D(a)/a$ of
the NADE model is $\sim 30\%$ larger than that of the $\Lambda$CDM
model. In the case of $n=3.5$ the relative difference is close to
$\sim 18\%$ and for $n=2.5$ we have $\sim 4\%$.

Comparing now the smooth and clustered NADE scenarios, we conclude
that at high $z$ the growth factor of the latter scenario is small
with respect to that of the former model, while prior to the present
epoch the growth factor of the smooth NADE model tends to that of
the clustered case.

\begin{figure}
 \centering
 \includegraphics[width=8cm]{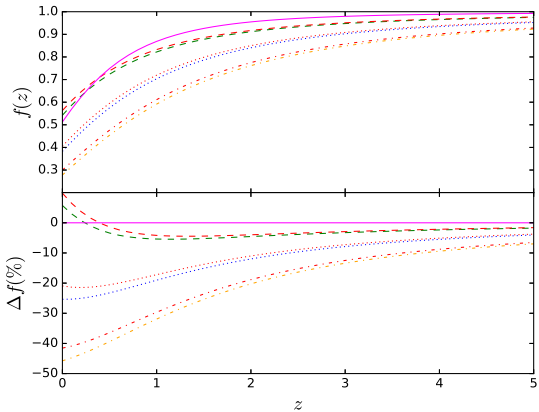}
 \caption{ The Redshift evolution of the growth rate of clustering
$f(z)$(top panel) and its fractional difference against $\Lambda$CDM
model $\rm \Delta f$ (bottom panel). Notice, that the lines are the
same to those of figure (\ref{figb}).}
 \label{figc}
\end{figure}

Lastly, we compute the growth rate of clustering in NADE
cosmologies. This quantity is defined as the logarithmic derivative
of $D$ with respect to $d{\rm ln}a$, $f=d\ln{D(a)}/d\ln{a}$. In
figure (\ref{figc}), we show the redshift evolution of $f(z)$ and
the fractional difference with respect to that of the concordance
$\Lambda$ model,$\Delta f(\%)=(f_{\rm NADE}-f_{\rm \Lambda})/f_{\rm
\Lambda}$. As expected, for the reasons developed above we have
$f_{\Lambda}(z) \simeq 1$ at high $z$, while in the case of NADE
models we find small but not negligible deviations from unity. We
also observe that in the redshift range $z\in [0,5]$ we have the
following results:

\begin{itemize}
\item  Smooth NADE model:
the relative deviation $\rm \Delta f$ lies in the interval
$\sim[5\%,-2\%]$ for $n=2.5$. While for $n=3.5$ and $n=4.5$ we find
$\sim[-25\%,-4\%]$ and $\sim[-45\%,-7\%]$ respectively.

\item  Clustered NADE model: here the relative difference is
$\sim [10\%,-2\%]$ for $n=2.5$. While for $n=3.5$ and $n=4.5$ we
find $\sim[-20\%,-4\%]$ and $[-40\%,-7\%]$ respectively.
\end{itemize}

We see that in the framework of clustered NADE the growth rate is
somewhat larger than the homogeneous case. We may understand this
feature based on the following arguments. In the homogeneous NADE
scenario the DE component is uniformly distributed both inside and
outside the cosmic structures. In other words DE acts against
gravity which implies that more suppression of the growth of matter
perturbations is taking place. While, if DE is allowed to clump in a
similar manner to dark matter then the total energy of DE can not
act against the force of gravity \citep[see
also][]{Malekjani:2015pza}.

\section{Observational constraints on the parameter of NADE model}
In this section we perform an overall likelihood analysis
using the latest cosmological data. Specifically, the total
likelihood function is the product of the individual
likelihoods:
\begin{equation}\label{eq:like-tot}
 {\cal L}_{\rm tot}({\bf p})={\cal L}_{\rm sn} \times {\cal L}_{\rm bao} \times {\cal L}_{\rm cmb} \times {\cal L}_{h} \times
 {\cal L}_{\rm \rm bbn} \times {\cal L}_{\rm gr}\;,
\end{equation}
so the total chi-square $\chi^2_{\rm tot}$ is given by:
\begin{equation}\label{eq:like-tot_chi}
 \chi^2_{\rm tot}({\bf p})=\chi^2_{\rm sn}+\chi^2_{\rm bao}+\chi^2_{\rm cmb}+\chi^2_{h}+\chi^2_{\rm bbn}+\chi^2_{\rm gr}\;.
\end{equation}
For more details regarding the statistical technique, the likelihood
functions, the cosmological data (see SNIa, BAO, CMB-shift
parameter, $H(z)$, Big Bang Nucleosythesis and growth rate data)
with the corresponding covariances, we refer the reader to
\citep{Basilakos:2009wi,Hinshaw:2012aka,Mehrabi:2015hva,Mehrabi:2015kta}
The vector ${\bf p}$ contains the free parameters of the particular
cosmological model. In the present analysis, the relevant parameters
are $(\Omega_{\rm DM},\Omega_b,\Omega_r,H_0,n,\sigma_8)$.
Considering a spatially flat universe, one can obtain $\Omega_{\rm
b}=1-\Omega_{d}-\Omega_{\rm DM}-\Omega_r$. Notice, that $\Omega_d$
is given by Eq. (\ref{omega}) and it only depends on the value of
$n$. The radiation density is fixed to $\Omega_r=2.469\times
10^{-5}h^{-2}(1.6903)$ \citep{Hinshaw:2012aka}.
%our free
%parameters
Therefore, if we use only the expansion data (SNIa, BAOs etc) then
the statistical vector becomes ${\bf p}=\{\Omega_{\rm DM},H_0,n\}$.
In the case of background and growth rate data we have ${\bf
p}=\{\Omega_{\rm DM},H_0,n,\sigma_8\}$. As far as the evolution of
the rms-variance at $R=8h^{-1}$Mpc is concerned we utilize the well
known formula
$\sigma_8(z)=\sigma_8\frac{\delta_m(z)}{\delta_{m}(z=0)}$. The next
step is to use the Markov Chain Monte Carlo (MCMC) procedure in
order to find the best fit values and their confidence regions [for
a similar analysis see \cite{Mehrabi:2015kta}].

\begin{figure*}
 \centering
 \includegraphics[width=.49\textwidth]{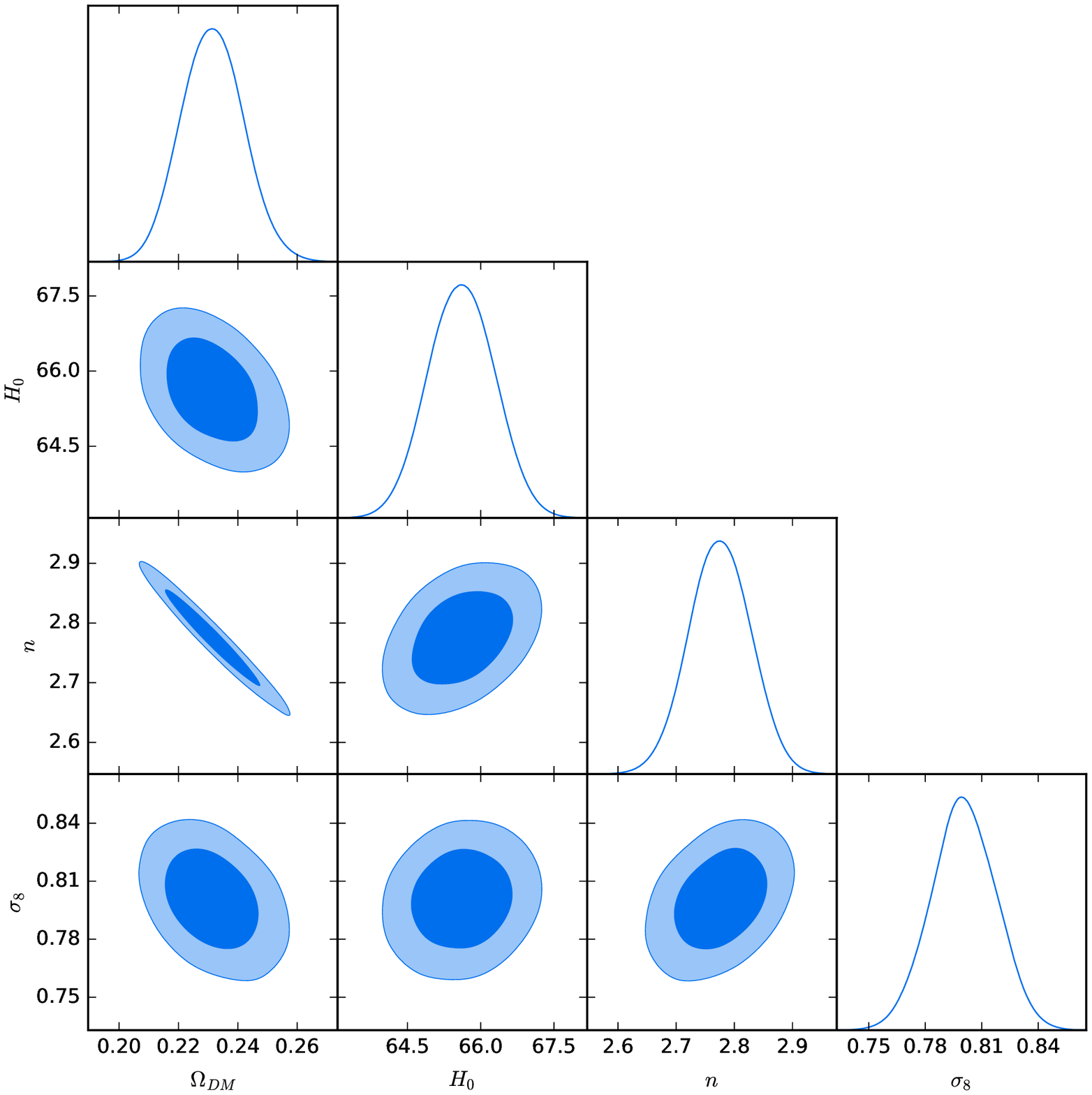} \includegraphics[width=.49\textwidth]{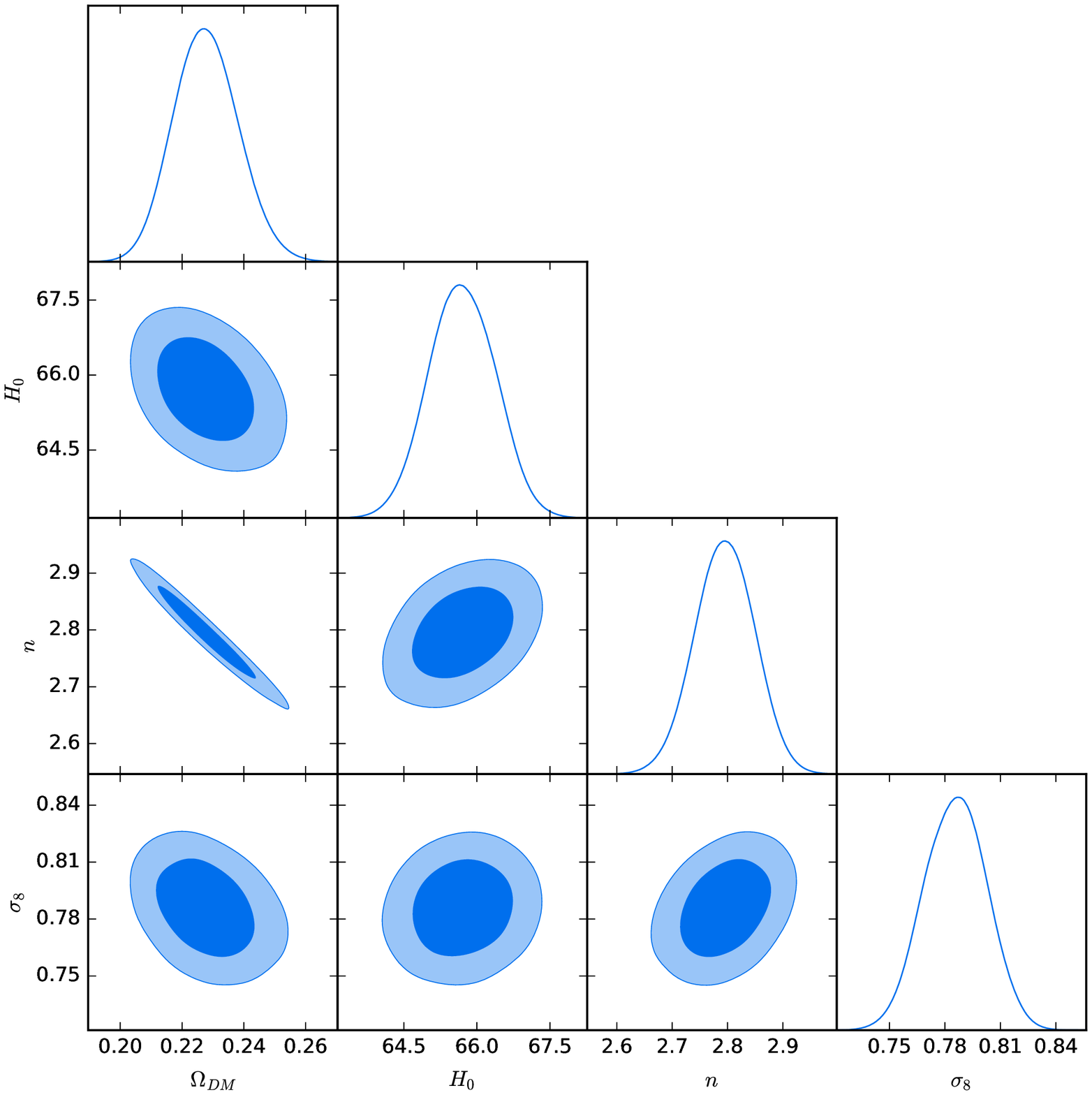}\\
 \includegraphics[width=.4\textwidth]{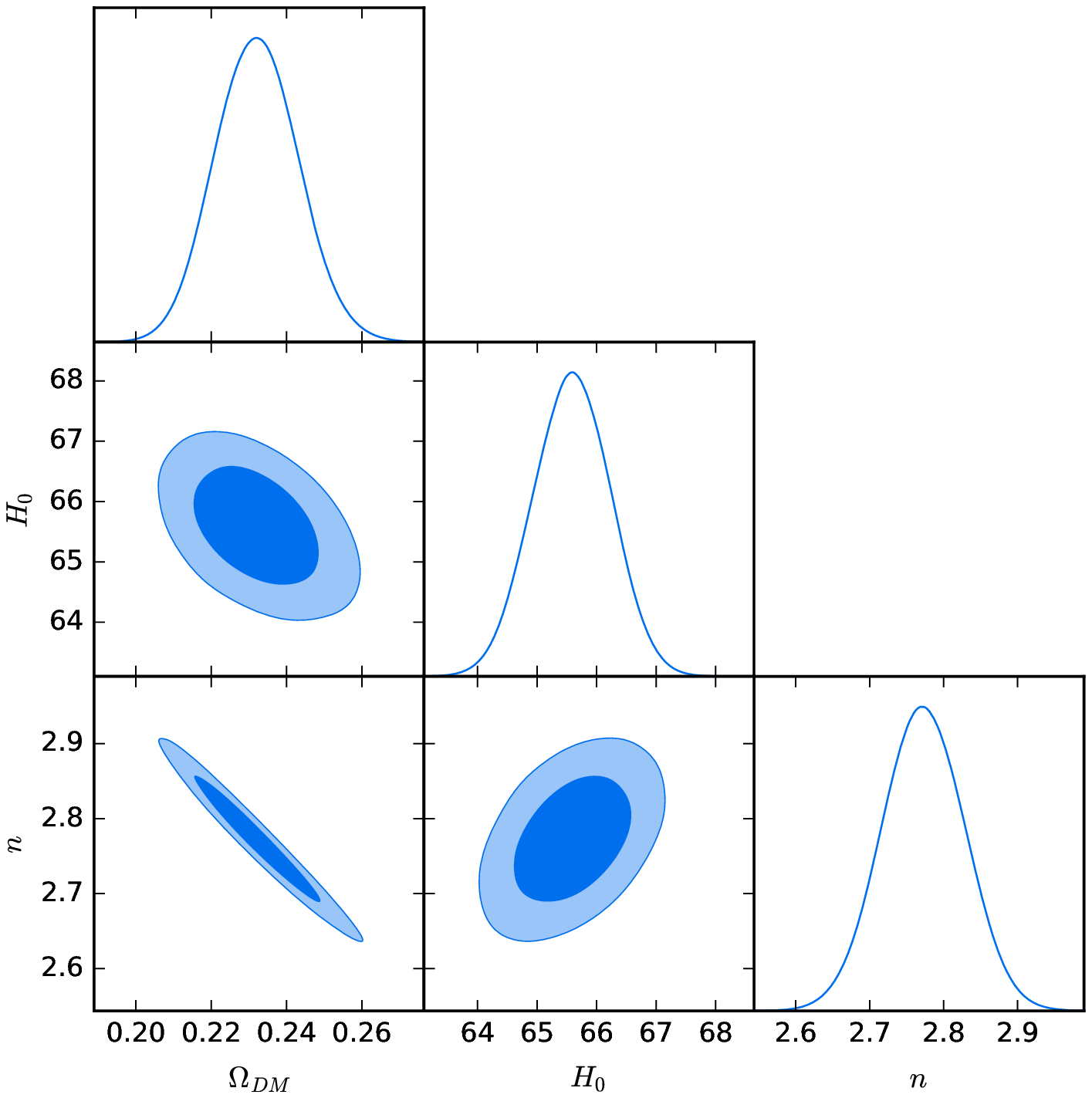}
 \caption{The overall (geometrical/growth)
likelihood contours ($1\sigma$ and $2\sigma$ confidence levels)
for various planes. Notice, that in the upper left (upper right) panel
we show the results for the
 smooth NADE (clustered NADE) model.
The bottom panel shows the results using only the background (geometrical) data
in the likelihood analysis.}
 \label{fig:like}
\end{figure*}

\begin{table}
 \centering
 \caption{The best fit values using only the expansion data.}
\begin{tabular}{c  c  c}
\hline \hline
 & NADE & $\Lambda$CDM\\
\hline $\Omega_{\rm DM}$ & $0.232\pm 0.011$ & $0.2333\pm 0.0098$\\
\hline
$ H_0 $& $ 65.59\pm 0.64 $ & $ 69.68\pm 0.73 $\\
\hline $ n $ & $2.773\pm 0.056$ &$ - $\\
\hline
 $ w_{d}(z=0) $ & -0.7966 &  -1.\\
\hline
 $ \Omega_{d}(z=0) $ & 0.7153 & 0.7208 \\
\hline \hline
\end{tabular}\label{tab:res_back}
\end{table}

\begin{table*}
 \centering
 \caption{The best fit values using  the expansion and the growth rate data.}
\begin{tabular}{c  c  c c}
\hline \hline
 & Smooth NADE & Clust. NADE & $\Lambda$CDM\\
\hline $\Omega_{\rm DM}$ & $0.232\pm 0.010$ & $0.228\pm 0.010
$&$0.2321\pm 0.0096$\\
\hline $ H_0 $& $65.61\pm 0.67$&$65.70\pm 0.68 $ &$69.73\pm 0.72
$\\
\hline
$ n $ & $2.775\pm 0.053$&$2.795\pm 0.054$ & $-$\\
\hline $  \sigma_{8} $ &$0.800\pm 0.017 $& $0.786\pm 0.017 $ &
$0.773\pm 0.017$\\
\hline
 $ w_{d}(z=0) $ & -0.7967 & -0.7971 & -1.\\
\hline
 $ \Omega_{d}(z=0) $ & 0.7157 & 0.7192 & 0.7218\\
\hline \hline
\end{tabular}\label{tab:res_all}
\end{table*}

Additionally, we implement the information criteria (IC) in order to
test the statistical performance of the models themselves.
Specifically, we use the BIC \citep{Schwarz:1974} and the AIC
\citep{Akaike:1974,Sugiura:1978} respectively. The BIC formula is
given by
\begin{equation}
{\rm BIC} = -2 \ln {\cal L}_{\rm max}+k\ln N
\end{equation}
where ${\cal L}_{\rm max}$ is the maximum likelihood, $k$ is the
number of parameters, and $N$ is the number of data points used in
the fit. Note that for Gaussian errors, $\chi^{2}_{min}=-2{\cal
L}_{\rm max}$, one can prove that the difference in BIC between two
models can be simplified to $\Delta BIC=\Delta \chi^{2}_{min}+\Delta
k \ln N$. In this context, owing to the fact that
$N/k \gg 1$ the AIC is defined as:
\begin{equation}
AIC = -2 \ln {\cal L}_{\rm max}+2k
\end{equation}
and thus $\Delta AIC=\Delta \chi^{2}_{min}+2\Delta k$.
Below we provide our statistical results.

In the case of the expansion data we find:\\
\begin{itemize}
\item  for the NADE model,  $\chi^{2}_{\rm min}=574.56$, AIC=580.56 and BIC=593.81;
\item for the $\Lambda$CDM model, $\chi^{2}_{\rm min}=575.03
$,  AIC=581.03 and BIC=594.28;
\end{itemize}

For the background and the growth rate data we have:,\\
\begin{itemize}
\item  homogeneous NADE model,  $\chi^{2}_{\rm min}=582.28$, AIC=590.28 and BIC=608.07;
\item  clustered NADE model,  $\chi^{2}_{\rm min}=583.05$, AIC=591.03 and BIC= 608.84;
\item concordance $\Lambda$CDM model, $\chi^{2}_{\rm min}=582.75$, AIC=590.75 and BIC=608.54.
\end{itemize}
The above results show that NADE (homogeneous or clustered) and
$\Lambda$CDM models fit the cosmological data equally well.
Moreover, in Tables (\ref{tab:res_back}) and (\ref{tab:res_all}),
one may see a more compact presentation of our constraints including
the estimated values of $w_{d}$, $\Omega_{d}(z=0)$ and $n$ (to be
used in our growth index analysis). In order to visualize the
solution space in Fig.(\ref{fig:like}) we present the $1\sigma$ and
$2\sigma$ confidence contours for various parameter pairs.

\begin{figure}
 \centering
 \includegraphics[width=8cm]{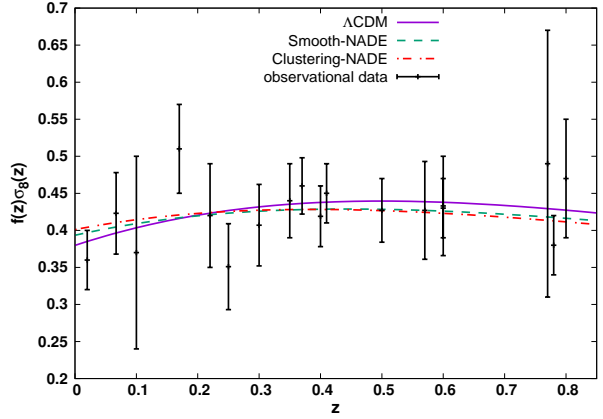}
 \caption{
Comparison of the observed and theoretical
evolution of the growth rate $f(z)\sigma_8(z)$.
The violet solid, green dashed and red
dotted-dashed curves correspond to $\Lambda$CDM, smooth NADE and clustered
NADE models. The open squares are
the growth data with their error bars.}
 \label{figf}
\end{figure}

In particular, using the background data we find $n=2.773\pm 0.056$.
Combining the background data and the growth data we obtain
$n=2.775\pm 0.053$ and $n=2.795\pm 0.054$ for smooth and clustered
NADE models respectively. $\Omega_{\rm DM}$ is negatively correlated
with the model parameter $n$ as expected from the fact that
$\Omega_{\rm d}$ in NADE model is correlated with $n^2$ by
definition. To this end, using the best fit values of Table
(\ref{tab:res_back}), we can compute the age of universe via the
following expression
\begin{equation}\label{cosmicage}
t_0=H_0^{-1}\int_{0}^{1}\frac{da}{aE(a)} \;.
\end{equation}
We find $t_{0}^{(\Lambda)}\simeq 13.80$ Gyr and $t_{0}^{(\rm
NADE)}\simeq 13.91$ Gyr which are in a good agreement the Planck
results ($13.81$ Gyr) \citep{Planck2015_XIII}. Concerning the
different values of $H_{0}$ between NADE and $\Lambda$CDM the
situation is as follows. Technically speaking, in order to obtain
the cosmic age of the universe via equation (\ref{cosmicage}), we
need to know a priori the Hubble constant and the corresponding
value of the integral. Based on the best fit cosmological parameters
(see Table \ref{tab:res_back}), we find that the numerical value of
the integral is 912.35 and 961.6 for NADE and $\Lambda$CDM
respectively. Therefore, a similar value of the cosmic age in NADE
and $\Lambda$CDM cosmological models requires that the Hubble
constant of the former scenario is small with respect to that of the
latter model. Finally, using the best fit cosmological parameters in
Fig.(\ref{figf}), we compare the theoretical evolution of the growth
rate $f(z)\sigma_8(z)$ with observational data. Notice, that
criteria of AIC and BIC indicate that both smooth and clustered NADE
models together with concordance $\Lambda$CDM model are consistent
with the growth data.

\section{Growth index in NADE model}
In this section we concentrate on the growth index of matter
fluctuations $\gamma$ which characterizes the growth rate of
clustering through the following formula \citep[first introduced
by][]{Peebles1993} \be \label{fzz221} f(z)=\frac{d{\rm ln}
\delta_{\rm m}}{d{\rm ln} a}(z)\simeq \Omega^{\gamma}_{m}(z) \;. \ee
In the literature there is a large body of studies which provides
the theoretical form of the growth index for various cosmologies,
including scalar field DE
\citep{Silveira:1994yq,Wang:1998gt,Linder:2003dr,Lue:2004rj,Linder:2007hg,Nesseris:2007pa},
DGP \citep{Linder:2007hg,Gong:2008fh,Wei:2008ig,Fu:2009nr},
Finsler-Randers \citep{Basilakos:2013ij}, running vacuum
$\Lambda(H)$ \citep{Basilakos:2015vra}, $f(R)$
\citep{Gannouji:2008wt,Tsujikawa:2009ku}, $f(T)$
\citep{Basilakos:2016xob} and Holographic dark energy
\citep{Mehrabi:2015kta}.

Now combining equations (\ref{pois}),
(\ref{dm},\ref{de},\ref{tm},\ref{td}) for $c_{\rm eff}^{2} \ne 0$
and using $\frac{d\delta_{\rm m}}{dt}=aH\frac{d\delta_{\rm m}}{da}$
we find \citep[see also][]{Abramo2007,Abramo:2008ip,Mehrabi:2015kta}
\begin{equation}\label{odedelta}
a^{2}\delta_{\rm m}^{\prime \prime}+
a\left(3+\frac{\dot{H}}{H^2}\right)\delta_{\rm m}^{\prime}=
\frac{3}{2}\Omega_{\rm m} \mu\;,
\end{equation}
where
\begin{equation}\label{eos22}
\frac{{\dot H}}{H^{2}}=\frac{d{\rm ln} H}{d{\rm ln}a}
=-\frac{3}{2}-\frac{3}{2}{\rm w}_{\rm d}(a)\Omega_{\rm d}(a)\;,
\end{equation}
and $\Omega_{\rm d}(a)=1-\Omega_{\rm m}(a)$. Notice, that $\mu(a)$
describes the properties of the NADE model, namely
\begin{equation} \label{VV}
\mu(a)=\left\{ \begin{array}{cc} 1
\;\;
       &\mbox{Homogeneous NADE}\\
  1+\frac{\Omega_{\rm d}(a)}{\Omega_{\rm m}(a)}\Delta_{\rm d}(a)(1+3c_{\rm eff}^2)
\;\;
       & \mbox{Clustered NADE}
       \end{array}
        \right.
\end{equation}
where $\Delta_{\rm d}\equiv \delta_{\rm d}/\delta_{\rm m}$. As
expected, if we impose $c_{\rm eff}^2=0$ then Eq.(\ref{odedelta})
boils down to Eq.(\ref{ddm}). Also, for the concordance $\Lambda$CDM
model we have $\delta_{\rm d}\equiv 0$ by definition.

Moreover, inserting Eq.(\ref{fzz221}) and Eq.(\ref{eos22}) in
Eq.(\ref{odedelta}) we obtain
\begin{equation}\label{Poll}
-(1+z)\frac{d\gamma}{dz}{\rm ln}(\Omega_{\rm m})+\Omega_{\rm m}^{\gamma}+
3{\rm w}_{\rm d}\Omega_{\rm d}\left(\gamma-\frac{1}{2}\right)+\frac{1}{2}=
\frac{3}{2}\Omega_{\rm m}^{1-\gamma} \mu \;.
\end{equation}
Concerning the evolution of the growth index one can use the
following well known approximation \citep[see
also][]{Polarski:2007rr,Wu:2009zy,Belloso:2011ms,DiPorto:2011jr,Ishak:2009qs,Basilakos:2012ws,Basilakos:2012uu}
\begin{equation}
\label{gzzz}
\gamma(a)=\gamma_{0}+\gamma_{1}\left[1-a(z)\right]\;.
\end{equation}
Therefore, using Eq.(\ref{Poll}) at the present time $z=0$ and
taking into account Eq.(\ref{gzzz}) we find \citep[see
also][]{Polarski:2007rr}
\begin{equation}
\label{Poll2}
\gamma_{1}=\frac{\Omega_{\rm m0}^{\gamma_{0}}+3{\rm w}_{d0}(\gamma_{0}-\frac{1}{2})
\Omega_{\rm d0}+\frac{1}{2}-\frac{3}{2}\Omega_{\rm m0}^{1-\gamma_{0}} \mu_{0}}
{{\rm ln}  \Omega_{\rm m0} }\;,
\end{equation}
where $\mu_{0}=\mu(z=0)$ and ${\rm w}_{d0}={\rm w}_{\rm d}(z=0)$.
Obviously, in order to treat the growth index evolution we need to
know the value of $\gamma_{0}$. This is given in terms of the
asymptotic value of $\gamma(z)$, namely $\gamma_{\infty}\simeq
\gamma_{0}+\gamma_{1}$ at high redshifts $z\gg 1$. The analytical
formula of $\gamma_{\infty}$ is given by \be \label{g000}
\gamma_{\infty}=\frac{3(M_{0}+M_{1})-2(H_{1}+N_{1})}{2+2X_{1}+3M_{0}}
\ee where the following quantities have been defined: \be
\label{Coef1} M_{0}=\left. \mu \right|_{\omega=0}\,, \ \
M_{1}=\left.\frac{d \mu}{d\omega}\right|_{\omega=0} \ee and \be
\label{Coef2} N_{1}=0\,,\ \ H_{1}=-\frac{X_{1}}{2}=\frac{3}{2}\left.
{\rm w_{\rm d}}(a)\right|_{\omega=0} \,. \ee For more details
regarding the above formula we refer the reader to
\cite{Steigerwald:2014ava} in which all the cosmological quantities
are given in terms of the variable $\omega={\rm ln}\Omega_{\rm
m}(a)$. This means that for $z\gg 1$ we have $\Omega_{\rm m}(a)\to
1$ [or $\Omega_{\rm d}(a)\to 0$] and thus $\omega \to 0$. As we have
already mentioned in section 2 at large enough redshifts in the
matter dominated era we have $\Omega_{d}\simeq n^{2}a^{2}/4$
\citep{Wei:2007xu} and thus $w_{\infty} \equiv w_{\rm d}(z\gg
1)\simeq -2/3$.

\begin{figure}
 \centering
 \includegraphics[width=8cm]{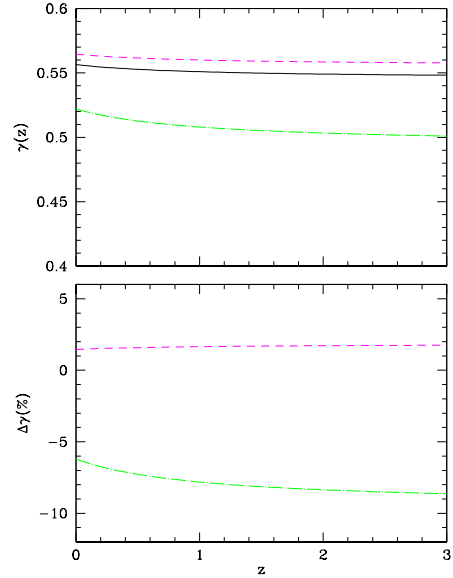}
 \caption{{\em Upper Panel:} The growth index versus redshift.
The dashed and the dotted dashed lines correspond to homogeneous
and clustered NADE models respectively. Notice, that
the solid line
corresponds to the usual $\Lambda$ cosmology.
{\em Bottom Panel:} The relative deviation
$[1-\gamma(z)/\gamma^{(\Lambda)}(z)]\%$ of the
NADE (homogeneous:dashed curve and clustered: dotted dashed curve)
models with respect to $\Lambda$CDM.}
 \label{growth}
\end{figure}

Let us now present our growth index results:
\begin{itemize}
  \item {\bf Homogeneous NADE model:} here we have
$\mu(a)=1$ ($\Delta_{\rm d}\equiv 0$).
Utilizing Eqs.(\ref{Coef1}) and (\ref{Coef2}) we obtain
$$
\{ M_{0},M_{1},H_{1},X_{1}\}=\{ 1,0,\frac{3{\rm w}_{\infty}}{2},-3{\rm w}_{\infty}\}
$$
and thus from Eq.(\ref{g000}) we find \be \gamma_{\infty}
=\frac{3({\rm w}_{\infty}-1)}{6{\rm w}_{\infty}-5} . \ee Obviously,
for $w_{\infty} \simeq -2/3$ we derive an asymptotic value
$\gamma_{\infty}\simeq 5/9$ which is close to that of the
$\Lambda$CDM model ($w_{\infty}=-1$),
$\gamma_{\infty}^{(\Lambda)}\simeq 6/11$. For comparison we also
provide the result of \citep{Mehrabi:2015kta} who found
$\gamma_{\infty}\simeq 4/7$ in the case of homogeneous HDE model.

Substituting now $\gamma_{0} \simeq \gamma_{\infty}-\gamma_{1}$ into
Eq.(\ref{Poll2}) and utilizing the cosmological constraints of Table
(\ref{tab:res_all}) we obtain $(\gamma_{0},\gamma_{1}) \simeq
(0.566,-0.01)$. For the $\Lambda$CDM model we have
$(\gamma_{0},\gamma_{1})^{(\Lambda)}\simeq (0.557,-0.012)$. In the
upper panel of Fig.(\ref{growth}) we show the growth index evolution
(\ref{gzzz}) for the homogeneous NADE (dashed curve) and $\Lambda$CDM
models (solid curve) respectively. We see that the evolution
of the growth index in homogeneous NADE model is somewhat larger than
the usual $\Lambda$CDM cosmological model. Specifically, as we can
see from the bottom panel of Fig.(\ref{growth}) the relative
difference $[1- \gamma^{(\rm NADE)}/ \gamma^{(\Lambda)}]$ is about
$\sim 1.7\%$.

\item {\bf Clustered NADE model:} In this case the quantity $\mu(a)$ is given
by the second branch of Eq.(\ref{VV}) which implies that we need to
treat the form of $\Delta_{\rm d}$. Based on Eq.(\ref{IC2}) we
arrive at \be \Delta_{\rm d}=\frac{1+{\rm w_{d}}}{1-3{\rm w_{d}}}=
\frac{1}{3}\frac{\sqrt{\Omega_{\rm d}}}{(2na-\sqrt{\Omega_{\rm
d}})}\;. \ee Within this context $\mu(a)$ becomes \be
\mu(a)=1+(1+3c_{\rm eff}^{2})\frac{\Omega_{\rm d}}{1-\Omega_{\rm d}}
\frac{1+w_{\rm d}}{1-3w_{\rm d}} \;. \ee Therefore, in the matter
dominated epoch $\Omega_{d}\simeq n^{2}a^{2}/4$ (or $w_{\rm d}\simeq
w_{\infty} \simeq -2/3$) we can easily obtain that $\Delta_{\rm d}
\simeq 1/9$ at high $z$ ($a\to 0$ or $\Omega_{\rm d} \to 0$). Thus
from Eqs.(\ref{Coef1}) and (\ref{Coef2}) we get
$$
\{ M_{0},M_{1},H_{1},X_{1}\}=\{ 1,-\frac{5(1+3c_{\rm eff}^2)}{27},\frac{3{\rm w}_{\infty}}{2},-3{\rm w}_{\infty}\}
$$
and from Eq.(\ref{g000}) we obtain
\be \label{g001}
\gamma_{\infty}\simeq \frac{40-15c_{\rm eff}^{2}}{81} \;.
\ee
In the
case of $c^{2}_{\rm eff}=0$ (fully clustered NADE model) we have
$\gamma_{\infty}\simeq 40/81\approx 1/2$ which is lower ($\sim 8\%$)
than the theoretically predicted value of the $\Lambda$CDM model
$\gamma^{(\Lambda)}_{\infty}\approx 6/11$. Also, we would like to
point out that our growth index
prediction is larger than that of the clustered HDE model, $\gamma_{\infty}
\approx 3/7$ \citep{Mehrabi:2015kta}.

Using the above value of $\gamma_{\infty}$, Eq.(\ref{Poll2}) and the
cosmological parameters of Table (\ref{tab:res_all}) we obtain
$(\gamma_{0},\gamma_{1})\simeq (0.522,-0.028)$. Finally, in the
upper panel of Fig(\ref{growth}) we show the evolution of
$\gamma(z)$ for the clustered NADE model (dotted dashed curve). In this
case, we observe that the growth index
deviates with respect to that
of the usual $\Lambda$ cosmology. Indeed, the relative deviation
lies in the interval $[-6.2\%,-8.6\%]$.

\end{itemize}

\section{Discussion and Conclusions}\label{conclude}

Combining the basic uncertainty principle in quantum mechanics
together with the gravitational effects of general relativity,
\cite{Cai2007} proposed a new model of DE the so-called agegraphic
dark energy model (ADE). Replacing the cosmic time $t$ with the
conformal time $\eta$, \cite{Wei:2007ty,Wei:2007xu} introduced in
the literature the new agegraphic dark energy (NADE) model in which
the corresponding EoS parameter is a function of redshift. The aim
of our paper is to investigate the main properties of the NADE model
at background and perturbation levels respectively.

The current study was performed by a three-step process.  Firstly,
solving the system of the main differential equations at the
background (Friedmann and continuity) and perturbation levels, we
investigated the behavior of the basic cosmological quantities
$\{H(z), w(z), \Omega_{\rm d}(z), D(z) \}$ in order to understand
the global characteristics of the NADE model (see Fig.\ref{figa}). We
verified that the EoS parameter remains in the quintessence regime
and it lies in the interval $-1 \leq w_{d} \leq -\frac{2}{3}$.
Notice, that for large values of $n$ the current value of the EoS
parameter tends  to -1. As expected, the cosmic expansion depends on
the choice of $n$. In particular, we found that for large values of
$n>3$, the parameter $\Omega_{\rm d}(z)$ of the NADE model strongly
deviates from that of $\Lambda$CDM. In this context we have that the
Hubble parameters obey: $H_{\rm NADE}(z)<H_{\Lambda}(z)$. Also, we
found that the growth factor $D_{\rm NADE}(z)$ seems to evolve at
high redshifts $z>2$ and the amplitude of $D_{\rm NADE} (z)$
increases as a function of $n$.
Note that at the same redshift range
$D_{\Lambda}(z)$ reached a plateau.
Potentially, the latter behavior
of $D(z)$ can be used to distinguish between NADE and $\Lambda$CDM
at perturbation level.

Secondly, we performed a joint statistical analysis, involving the
latest geometrical data (SNe type Ia, CMB shift parameter and BAO
etc) and growth data and found that the combined statistical
analysis, within the context of flat FRW space, can place tight
constrains on the main cosmological parameters giving the reader the
opportunity to appreciate the precision of our statistical results.
In particular, we found $n=2.775\pm 0.053$ ($\sigma_{8}=0.800\pm
0.017$) and $n=2.795\pm 0.054$ ($\sigma_{8}=0.786\pm 0.017$)
  for homogeneous and inhomogeneous (DE is allowed to clump)
NADE models respectively. Notice, that the present value of
$\Omega_m$ is close to 0.29 in all cases. It is interesting to
mention that the above constraints are in agreement with those of
\cite{Wei:2007xu} \citep[see also][]{Wei:2008rv,Zhang:2012pr}. Using
the aforementioned cosmological parameters we found that the age of
the universe is $t_{0}^{(\rm NADE)}\simeq 13.91$ Gyr which differs
from that of the Planck \citep{Planck2015_XIII} by $\sim 1\%$.
Lastly, using the basic information criteria (AIC and BIC), we
concluded that both smooth and clustered NADE scenarios as well as
the $\Lambda$CDM model fit the observationally data equally well.

Thirdly, we studied the performance of NADE model at the
perturbation level. Specifically, following the methodology of
\citep{Steigerwald:2014ava} we estimated for the NADE model the
asymptotic value of the growth index of linear matter fluctuations
($\gamma$). Considering a homogeneous NADE model we found $\gamma
\approx 5/9$ which is close to that of the traditional $\Lambda$
cosmology $\gamma^{(\Lambda)} \approx 6/11$. On the other hand, if
we allow clustering in NADE then we obtained $\gamma \approx 1/2$,
which is $\sim 8\%$ smaller than that of the $\Lambda$CDM model.
Finally, we extended the growth analysis in the case where $\gamma$
varies with redshift and found that the $\gamma(z)$ is quite large
with respect to that of the clustered NADE scenario. This implies
that the clustered DE scenario can be differentiated from the other
two models on the basis of the growth index evolution.

\bibliographystyle{mnras}
\bibliography{ref}

\label{lastpage}

\end{document}